\documentstyle[psfig,pstricks,pst-plot]{article}

\def\r{{\rm\bf r}}

\newcommand{\mc}[1]{\multicolumn{2}{|c|}{#1}}
\newcommand{\rb}[1]{\raisebox{1.5ex}[-1.5ex]{#1}}
\def\r{{\bf{r}}}
\def\R{{\bf{R}}}
\def\G{{\bf{G}}}

\def\k{{\bf{k}}}
\def\K{{\bf{K}}}

\def\ba{\begin{array}}
 \def\ea{\end{array}}
\def\bc{\begin{center}}
 \def\ec{\end{center}}
\def\be{\begin{equation}}
 \def\ee{\end{equation}}
\def\bea{\begin{eqnarray}}
 \def\eea{\end{eqnarray}}
\def\bi{\begin{itemize}}
 \def\ei{\end{itemize}}
\def\bn{\begin{enumerate}}
 \def\en{\end{enumerate}}
\def\bF{\begin{figure}}
 \def\eF{\end{figure}}

\def\bt{\begin{tabular}}
 \def\et{\end{tabular}}
\def\bT{\begin{table}}
 \def\eT{\end{table}}

\def\ed{\end{document}}
\def\>{$\rightarrow$}
\def\=>{$\Rightarrow$}

\def\nn\\{\nonumber \\}

\def\ACM#1{ACM Trans.\,Math.\,Softw.\,{\bf #1}}

\def\AQC#1{Advances in Quantum Chemistry {\bf #1}}

\def\CPC#1{Comp.\,Phys.\,Commun.\,{\bf #1}}

\def\IJQCS#1{Int.\,J. Quantum.\,Chem. Suppl.\,{\bf #1}}

\def\JPCS#1{J.\,Phys.\,Chem.\,Solids\,{\bf #1}}
\def\JPF#1{J.\,Phys.\,F\,{\bf #1}}

\def\MCP#1{Meth.\,Comp.\,Phys\,{\bf #1}}

\def\PKM#1{Phys.\,Kondens.\,Mater.\,{\bf #1}}

\def\PR#1{Phys.\,Rev.\,{\bf #1}}

\def\PRB#1{Phys.\,Rev.\,B~{\bf #1}}
\def\PRL#1{Phys.\,Rev.\,Lett.\,{\bf #1}}

\def\SSC#1{Solid State Commun.\,{\bf #1}}
\def\SSP#1{Solid State Phys.\,{\bf #1}}

\def\ZPB#1{Z.\,Phys.\,B~{\bf #1}}

\begin{document}
\bc
{\Large\bf Improving the Efficiency of FP-LAPW Calculations}\\[1.cm]
{\large Max Petersen, Frank Wagner, Lars Hufnagel, and 
Matthias Scheffler\\
\small\em Fritz-Haber-Institut der Max-Planck-Gesellschaft,\\[-.2cm]
\small\em Faradayweg 4-6, D-14\,195 Berlin (Dahlem), Germany\\[.5cm]
\large Peter Blaha, and Karlheinz Schwarz\\
\small\em Institut f. Physikalische und Theoretische Chemie,\\[-.2cm]
\small\em Technische Universit\"at Wien\\[-.2cm]
\small\em Getreidemarkt 9/156, A-1060 Vienna, Austria
}\ec

\begin{abstract}{

The {\em full-potential linearized augmented-plane wave} (FP-LAPW)
method is well known to enable most accurate calculations of the
electronic structure and magnetic properties of crystals and surfaces.
The implementation of atomic forces has greatly increased it´s
applicability, but it is still generally believed that FP-LAPW
calculations require substantial higher computational effort compared
to the pseudopotential plane wave (PPW) based methods.

In the present paper we analyse the FP-LAPW method from a
computational point of view. Starting from an existing implementation
(WIEN95 code), we identified the time consuming parts and show how
some of them can be formulated more efficiently. In this context also
the hardware architecture plays a crucial role.  The remaining
computational effort is mainly determined by the setup and
diagonalization of the Hamiltonian matrix. For the latter, two
different iterative schemes are compared. The speed-up gained by these
optimizations is compared to the runtime of the ``original'' version
of the code, and the PPW approach.  We expect that the strategies
described here, can also be used to speed up other computer codes,
where similar tasks must be performed.

}

\vspace{.5cm}

\noindent
\begin{tabular}{l}
\hline
\end{tabular}
\end{abstract}

\twocolumn
\newpage
\section*{PROGRAM SUMMARY}
{\footnotesize\sloppy\baselineskip 10pt
{\em Title of program extension:} \verb|wien-speedup|\\[1ex]
{\em catalogue number:} ... \\[1ex]
{\em Program obtainable from:}
 CPC Program Library, Queen's University of Belfast, N. Ireland
(see application form in this issue)\\[1ex]
{\em CPC Program Library programs used:}
{\em cat. no.:} ABRE;
{\em title:}    \verb|WIEN|;
{\em ref. in CPC:} 59 (1990) 399\\[1ex]
{\em Licensing provisions:} none \\[1ex]
{\em Computer, operating system, and installation:}
\bi
\item IBM RS/6000; AIX; Fritz-Haber-Institut
 der Max-Planck-Gesellschaft; Berlin.
\ei
{\em Operating system:} UNIX\\[1ex]
{\em Programming language:} FORTRAN77\\
(non-standard feature is the use of \verb|ENDDO|)\\[1ex]
{\em floating point arithmetic:} 64\,bits\\[1ex]
{\em Memory required to execute with typical data:}\\
64\,Mbyte (depends on case)\\[1ex]
{\em No.~of bits in a word:} 64\\[1ex]
{\em No.~of processors used:} one\\[1ex]
{\em Has the code been vectorized?} no\\[1ex]
{\em Memory required for test run:} ~64\,MByte\\[1ex]
{\em Keywords} \\
density-functional theory, linearized augmented plane wave
method, LAPW, supercell, total energy, forces, structure
optimization, molecular dynamics, crystals, surfaces,
molecules\\[1ex]
{\em Nature of the physical problem}\\
For {\it ab-initio} studies of the electronic and
magnetic properties of poly-atomic systems, such
as molecules, crystals, and surfaces. 
\\[1ex]
{\em Method of solution}\\
The full-potential linearized augmented plane wave (FP-LAPW)
method is well known to enable accurate calculations of
the electronic structure and magnetic properties of crystals
\cite{slat,bros,louc,SkoelXX,SandeXX,Swimm81,Sjans84,Smatt86,Sblah90,Sblah93a,Ssing94}.
Within the supercell approach it has also been used for
studies of defects in the bulk and for crystal surfaces.
}

{\renewcommand{\,}{$\!$\ }
\renewcommand{\refname}{\ }
{\em References}\\
}

\onecolumn

\section*{LONG WRITE-UP}
\section{Introduction}

\normalsize
The augmented plane wave (APW)
method~\cite{slat37,slat64,louc67,matt68,dimm71} and in particular its
linearized form, the LAPW
method~\cite{brosXX,marc67,koelXX,andeXX,wimm81,jans84,matt86,blah90,sing94},
enables accurate calculations of electronic and magnetic properties of
poly-atomic systems using density-functional theory (DFT)
~\cite{hohkohn,konsham}.  One successful implementation of the
full-potential LAPW (FP-LAPW) method is the program package
\verb|WIEN|, a code developed by Blaha, Schwarz and
coworkers~\cite{blah90}. It has been successfully applied to a wide
range of problems such as electric field gradients
~\cite{blah85,dufek95} and systems such as high-temperature
superconductors~\cite{schwarz90}, minerals~\cite{winkler96}, surfaces
of transition metals ~\cite{kohl95}, or anti-ferromagnetic
oxides~\cite{Wang} and even molecules~\cite{kohl96}.  Minimizing the
total energy of a system by relaxing the atomic coordinates for
complex systems became possible by the implementation of atomic
forces~\cite{kohl96}, and even molecular dynamics became feasible.  Up
to now the main drawback of the FP-LAPW-method compared to the
pseudopotential plane-wave (PPW) (e.g. Ref.~\cite{fhi96md} and
references therein) approach has been its higher computational
expense.  This may be mainly due to a discrepancy in optimization
efforts spent on both methods, and therefore we have analysed the
FP-LAPW method from a computational/numerical point of view. Starting
from the \verb|WIEN95| implementation~\cite{blah95}, we identified the
time consuming parts and will show how some of them can be formulated
more efficiently. In this context also the influence of the underlying
hardware architecture will be discussed.\\ The remainder of the paper
is organized as follows.  After introducing the principles of DFT and
summarizing the concepts of the FP-LAPW-method (Section \ref{DFT} and
Section \ref{lapwmethod}), we will report on our improvements made on
the \verb|WIEN95| implementation of the FP-LAPW-method (Section
\ref{opt}).  In Section \ref{results} we will show, how these
improvements make the FP-LAPW-method a strong competitor to the
popular PPW approach by comparing the run-times necessary to converge
a 9 layer slab of (4$\times$2)-Cu\,(110) ({\em i.e.} 72 atoms and 792
valence electrons) using both methods.

\section{Density-Functional Theory}\label{DFT}

The central statement of DFT is, that the problem of finding the
ground-state energy of a many-particle system, characterized by a
many-particle wavefunction $\Psi_0$, can be mapped on a
physically equivalent problem of finding the ground-state electron
density $n_0$, {\em i.e.}
\begin{eqnarray}
E[\Psi_0] & = & E[n_0] \quad {\rm with} \label{hohkohn1}\\
n_0(\r) & = & 
\langle\,\Psi_0\,|\,
\sum_{\alpha}^N \delta ( \r - \r_{\alpha} )\,|\, \Psi_0\,\rangle 
\quad ,
\end{eqnarray}
where $\r_{\alpha}$ is the coordinate of the $\alpha$-th electron.
The central statement of the Hohenberg-Kohn theorem~\cite{hohkohn} is,
that for an $N$ electron system the functional $E[n]$ is minimized 
by the ground-state electron density, $n_0$.
\begin{eqnarray}
E[n_0] & = & {\rm Min} \, E[n] \quad {\rm with\,\,the\,\,constraint} 
\label{hoko1} \\
\int n d^3r & = & N \quad . \label{hoko2}
\end{eqnarray}
In the Kohn-Sham formulation the functional $E[n]$ is split into the
following terms:
\begin{eqnarray}
E[n] & = & T_{\rm s}[n] + U[n] + E_{\rm xc}[n]\label{hohkohn2} \quad ,
\end{eqnarray}
the kinetic energy functional of non-interacting particles, $T_{\rm
s}[n]$, the functional of the electrostatic energy, $U[n]$ and the
rest, called exchange-correlation energy, $E_{\rm xc}$.  With
Eq.~(\ref{hohkohn2}), {\em i.e.} with the introduction of the
functional $T_{\rm s}[n]$, the variational problem of
Eqs.~(\ref{hoko1}),(\ref{hoko2}) becomes becomes equivalent to the
problem of solving a system of single-particle equations, called the
Kohn-Sham equations~\cite{konsham},
\begin{eqnarray}
h\varphi_i & = & \left[-\frac{\hbar^2}{2m_{\rm e}}{\boldmath \nabla^2} + 
V_{\rm eff}\right]\varphi_i = \epsilon_i\varphi_i 
\label{konsham} \\
n & = & \sum_i f_i \varphi_i^*\varphi_i \quad \label{dens} . 
\end{eqnarray}
Here, $-\frac{\hbar^2}{2m_{\rm e}}{\boldmath \nabla^2}$ is the
single-particle kinetic energy operator and $V_{\rm eff}$ is the
potential defined by the functional derivative of $U[n] + E_{\rm
xc}[n]$, 
\begin{eqnarray}
V_{\rm eff} & = & \frac{\delta (U + E_{\rm xc})}{\delta n}\quad .
\label{konsham2}
\end{eqnarray}
The electron density is obtained from Eq.~(\ref{dens}), where $f_i$
are the occupation numbers given by the Fermi distribution.  In
practice Eqs.~(\ref{konsham}), (\ref{dens}) and (\ref{konsham2}) are
solved in a selfconsistent field (SCF)-cycle: {\em i.e.} starting with
density $n_1$ one calculates the potential $V_{\rm eff}$, solves
Eq.~(\ref{konsham}) and by evaluating Eq.~(\ref{dens}) one obtains the
new density $n_2$, which leads to the next iteration cycle.

\section{The FP-LAPW-Method}
\label{lapwmethod}
In the augmented plane-wave (APW) method space is divided into an
interstitial region (IR) and non-overlapping muffin-tin (MT) spheres
centered at the atomic sites \cite{slat37}.  This allows an accurate
description of both, the rapidly changing (oscillating) wavefunctions,
potential and electron density close to the nuclei as well as the
smoother part of these quantities in between the atoms.  In the IR the
basis set consists of plane waves $\exp(i\K\cdot\r)$.  The choice of a
computationally efficient and accurate representation of the
wavefunctions within the MT spheres has been discussed by several
authors, e.g. \cite{louc67,brosXX,marc67,andeXX}.  In the original APW
formulation introduced by Slater \cite{slat37,slat64}, the plane-waves
are augmented to the exact solutions of the Schr\"odinger equation
within the MT at the calculated eigenvalues.  This approach is
computationally expensive because it leads to an explicit energy
dependence of the basis functions (and consequently of the Hamilton-
and overlap-matrices) and thus to a non-linear eigenvalue problem.
Instead of performing a single diagonalization to solve the KS
equation one repeatedly needs to evaluate (for many trial energies)
the determinant of the secular equation in order to find its zeros and
thus the single particle eigenvalues $\epsilon_i$. Going into the
complex energy plane would have been one option but was not explored
so far, except in an other context (see e.g. \cite{bormet} and
references therein).

In the linearized APW method the problem of the energy dependence of
the basis set is removed by using a fixed set of suitable MT radial
functions \cite{brosXX,marc67,andeXX}.  Within Andersen's approach,
used also in the \verb|WIEN| code, inside each atomic sphere $I$ and
for azimuthal quantum number $l$ the radial solutions
$u_l^I(\epsilon_l^I,r_I)$ of the KS equation at fixed energies
$\epsilon_l^I$ and their energy derivatives
$\dot{u}_l^I(\epsilon_l^I,r_I)$ are used as basis functions.
Basically, this choice corresponds to a linearization of the energy
dependence of $u_l^I(\epsilon,\r)$ around $\epsilon_l^I$
\cite{andeXX}.  The concept implies that the radial functions
$u_l^I(\epsilon_l)$ and $\dot{u}_l^I(\epsilon_l)$ and the respective
overlap and Hamilton matrix elements need to be calculated only for a
few energies $\epsilon_l^I$.  Moreover, all KS energies $\epsilon_i$
are found, for each ${\rm\bf k}$-point, by only one diagonalization
(for a detailed discussion see \cite{sing94}).

The LAPW basis functions $\phi_{\G}(\r,\k)$ which are used for the
expansion of the KS wavefunctions
\be
\psi_i(\r,\k) = \sum_{|\k+\G|\leq G^{\rm wf}} c_{i}(\k+\G)\, \phi_\G(\r,\k)
\label{KSexp}
\ee
are defined as
\be\label{wfbas}
\phi_{\G}(\r,\k) =
\left\{ \ba{ll}
\Omega^{-1/2} \exp(i(\k+\G)\cdot\r),& \r \in {\rm IR} \\
{\displaystyle \sum_I \sum_{lm}[
a_{lm}^I(\k+\G)\,u_l^I(\epsilon_l^I,r_I) + 
b_{lm}^I(\k+\G)\,\dot{u}_l^I(\epsilon_l^I,r_I)
]Y_{lm}(\hat{\r}_I)},& r_I \le s_I \quad.
\ea\right.
\ee
Here, $\G$ denote the reciprocal lattice vectors
and $\k$ a vector within the first Brillouin zone.
The wave function cutoff $G^{\rm wf}$ limits the number of
the $\G$ vectors and thus the size of the basis set.
The symbols in Eq.\,(\ref{wfbas}) have the following meaning:
$\Omega$ is the unit cell volume, $s_I$ is the MT radius, and
$\r_I = \r-\R_I$ is a vector within the MT sphere of the $I$-th atom.
Note that $Y_{lm}(\hat{\r})$ represents a complex spherical
harmonic with $Y_{l-m}(\hat{\r})=(-1)^mY^*_{lm}(\hat{\r})$.
The radial functions $u_l(\epsilon_l,r)$ and
$\dot{u}_l(\epsilon_l,r)$ are solutions of the equations
\bea\label{urad}
H^{\rm sph}\,u_l(\epsilon_l,r) &=&
\epsilon_l\, u_l(\epsilon_l,r)   \\
\nopagebreak
\label{udot}
H^{\rm sph}\,\dot{u}_l(\epsilon_l,r)  &=&
 [\epsilon_l \dot{u}_l(\epsilon_l,r) +
u_l(\epsilon_l,r)] 
\eea
which are regular at the origin. The operator $H^{\rm sph}$ contains
only the spherical average, {\em i.e.}, the $l=0$ component, of the
effective potential within the MT.  The $\epsilon_l$ should be chosen
near the center of the energy band with the corresponding $l$-character.  The
coefficients $a_{lm}(\k+\G)$ and $b_{lm}(\k+\G)$ are determined by requiring
that value and slope of the basis functions are continuous at the
surface of the MT sphere

The representation of the potential and electron density resembles
the one employed for the wave functions, {\em i.e.},
\be\label{vbas}
n^{\rm eff} (\r)=
\left\{\ba{ll}
\displaystyle\hspace*{3.5mm}
\sum_I \sum_{lm} n^{\rm eff}_{lm,I}(r_I)\, Y_{lm}(\hat{\r}_I),&
r_I \le s_I\\
\displaystyle
\sum_{|\G|\leq G^{\rm pot}} n_\G^{\rm eff}\, \exp(i\G\cdot\r),
& \r \in {\rm IR} \quad.
\ea\right.
\ee
Thus, no shape approximation is introduced and therefore such an 
approach is called
a full-potential treatment. The quality of this
description is controlled by the cutoff parameter
$G^{\rm pot}$ for the lattice vectors $\G$ and the number of the
$(l,m)$-terms included inside the MTs.

\section{Improving the {\tt WIEN} code}
\label{opt}

\subsection{Optimization strategies}
\label{opt_strat}

To achieve an optimal performance of a computer code on modern
computers, it is essential that the used algorithms match the
underlying hardware architecture.  On todays computers, often the
memory bandwidth is the limiting factor, {\em i.e.} the floating-point
operation units are stalled, waiting for data.  Then the performance
is not determined by the number of floating point operations per
second, but by the necessary number of load/store
operations. Therefore a significant objective of optimizing a code is
to reduce the communication between the processor and the relatively
slow memory, but to make optimum use of the fast cache. Thus the well
known fact for parallel computers, that an efficient use of
communication is crucial for complex and time consuming calculations,
holds also on stand-alone workstations.  The best way to improve the
performance of a program on a wide range of architectures without
loosing portability, is to write the code in such a way that the bulk
of the calculations is performed by calls to the well known {\sl basic
linear algebra subprograms (BLAS)} \cite{blas1,blas1b}; efficency can
then be obtained by using optimized implementations of these routines,
specifically tailored to the hardware used.  While on vector machines,
the so called Level 2 {\sl BLAS} routines (matrix-vector-operations)
lead to very satisfactory results, this approach is often not well
suited for architectures of modern high-performance workstations or
shared memory systems with a hierarchy of memory (registers, cache,
local memory, swap space).  For those architectures it is preferable
to partition the matrices into blocks and to perform the computation
by matrix-matrix operations on these blocks. This leads to a full
reuse of data already held in cache (or local memory) and reduces data
movement. While for Level 1 (vector-vector-operations) and Level 2
{\sl BLAS} routines the number of load/store operations is
proportional to the number of floating-point operations, the Level 3
(matrix-matrix-operations) approach \cite{blas3} gives a
surface-to-volume effect, {\em i.e.} if the matrices are of order $n$,
the number of floating-point operations is of order $n^3$, while the
number of load/store operations is of order $n^2$.  This minimizes the
influence of a limited memory bandwidth on the performance of the
program.  Therefore the goal in optimizing the code must be to use
Level 3 {\sl BLAS} routines as much as possible.

\subsection{The structure of the {\tt WIEN}-code}
The SCF cycle of the \verb|WIEN| code consist of five independent 
programs:
\begin{enumerate}
\item {\tt LAPW0:} generates the potential from a given charge density
\item {\tt LAPW1:} computes the eigenvalues and eigenvectors
\item {\tt LAPW2:} computes the valence charge density from the eigenvectors
\item {\tt CORE:} computes the core states and densities
\item {\tt MIXER:} mixes the densities generated by {\tt LAPW2} and {\tt CORE} with
the
density of the previous iteration to generate a new charge density
\end{enumerate}

From these programs {\tt LAPW1} and {\tt LAPW2} are the most time
consuming, while the time needed to run {\tt CORE} and {\tt MIXER} are
basically negligible.  Further inspection showed, that for example on
{\sl IBM RS/6000} nodes the performance of {\tt LAPW2} was far below
the theoretical peak performance, which indicates a poor adaptation of
the code to this hardware architecture. The optimizations done on {\tt
LAPW2} are described in Section \ref{lapw2}.  The situation was
different in the case of {\tt LAPW1}. Due to the use of standard
library routines the diagonalization of the matrix, which is the most
time consuming part, performes quite well on {\sl IBM RS/6000}
nodes. However, on several other hardware platforms with
substantially slower memory bandwidth, the performance was not so good
and those routines were also modified to increase performance.  Thus
further improvement on {\sl IBM RS/6000} could only be reached by
implementing a new algorithm.  Based on the fact that the matrix to be
diagonalized changes only little from iteration to iteration during
the selfconsistency cycle, an iterative diagonalization scheme could
be an attractive alternative.  We implemented two such schemes, which
use the information from the previous step to speed up the
diagonalization.  The details will be described in Section
\ref{lapw1}.

\subsection{{\tt LAPW2}: Generating the electron density}
\label{lapw2}
In {\tt LAPW2} the eigenvalues and eigenvectors found by {\tt LAPW1}
are read in.  The $\k$-space integration over the Brillouin zone (BZ)
is replaced by a finite $\k$-summation, in which each $\k$-point
contributes with a weight, $W_j(\k)$, in which for convenience also
the occupation factor of state $\epsilon_j$ ({\em i.e.} the
Fermifactor) is stored.  First the Fermi energy and then the expansion
of the valence electron density is calculated for each of the occupied
states at all $\k$-points in the irreducible part of the BZ.  The
valence electron density consists of two types of components: the electron
density inside each sphere $I$, $n^I({\bf r})$, represented in spherical
harmonics on a radial grid and the interstitial electron density,
$n^{IR}({\bf r})$ expressed as Fourier series.

\subsubsection{The electron density inside the MT-spheres}

The valence electron density inside a sphere is given by the expression:

\begin{eqnarray}
n^I({\bf r})
&=&\sum_{l''m''} n^{\rm eff}_{l''m'',I}(r_I)\, Y_{LM}(\hat{\r}_I) \quad
r_I \le s_I\\
  &=&\sum_{\k,j}W_j(\k)\sum_{lm}\sum_{l'm'}\sum_{\bf G,G'}\nonumber\\
               && \left\{ 
   c^*(j,\k+\G)\,a^{I*}_{lm}({\k+\G})\,u_l({r})\,
   c(j,\k+\G') \,a^I_{l'm'} ({\k+\G'})\,u_l'({r})
                \right.\nonumber\nopagebreak\\
    &&+ c^*(j,\k+\G)\,b^{I*}_{lm}({\k+\G})\dot{u}_l({r})\,
c(j,\k+\G')\,a^I_{l'm'}({\k+\G'})\,u_l'({r})
          \nonumber\nopagebreak\\
     &&+ c^*(j,\k+\G)\,a^{I*}_{lm}({\k+\G})\,u_l({r})\,
c(j,\k+\G')\,b^I_{l'm'}({\k+\G}')\dot{u}_l'({r})
          \nonumber\nopagebreak\\
 &&+ c^*(j,\k+\G)\,b^{I*}_{lm}({\k+\G})\dot{u}_l({r})\,
 c(j,\k+\G')\,b^I_{l'm'}({\k+\G'})\dot{u}_l'({r})
            \left.
               \right\}Y^*_{lm}(\hat{\r})Y_{l'm'}(\hat{\r}) \quad .
\end{eqnarray}

With the definition

\begin{eqnarray}
A^I_{lmj}(\k)&:=&
\sum_{\G} c(j,\k+\G)\, a^I_{lm}({\k+\G}) \label{alm}\\
B^I_{lmj}(\k)&:=& 
\sum_{\G} c(j,\k+\G)\, b^I_{lm}({\k+\G})
\label{blm}
\end{eqnarray}

the electron density reads:

\begin{eqnarray}
n^I({\bf r})
             &=&\sum_{\k,j}W_j(\k)\sum_{lm}\sum_{l'm'}
                \left\{ 
              A^{I *}_{lmj}(\k)A^{I}_{l'm'j}(\k)u_l({r})u_{l'}({r})
                \right.\nonumber\\
    &&+ B^{I*}_{lmj}(\k)A^{I}_{l'm'j}(\k)\dot{u}_l({r})u_{l'}({r})
%          \nonumber\\
     + A^{I*}_{lmj}(\k)B^{I}_{l'm'j}(\k)u_l({r})\dot{u}_{l'}({r})
          \nonumber\\
 &&+ B^{I*}_{lmj}(\k)B^{I}_{l'm'j}(\k)\dot{u}_l({r})\dot{u}_{l'}({r})
            \left.
               \right\}Y^*_{lm}(\hat{r})Y_{l'm'}(\hat{\r}) \quad .
\end{eqnarray}

It is obvious that the calculation of the sums (\ref{alm}),
(\ref{blm}) which run over all {\bf G}-vectors for every combination
of $(I,j,lm)$, will be the most time consuming part, and thus needs
special care to implement it efficiently.  The
straight forward implementation of the summation, as done in the
original \verb|WIEN| code, results in a high ratio of load/store
operations per floating-point operation and a very poor
performance.  A closer look shows that these formulas can be rewritten
in the form of a matrix-matrix-multiplication:

\begin{eqnarray}
A^{I,\k}(j,lm) &=& \sum_{\G} c(j,\k+\G)\, a^I({\k+\G},lm)\\
B^{I,\k}(j,lm) &=& \sum_{\G} c(j,\k+\G)\, b^I({\k+\G},lm)
\quad .
\end{eqnarray}

In this way the matrices $A^{I,\k}(j,lm)$ and $B^{I,\k}(j,lm)$ can be
calculated using optimized {\sl (BLAS-3)} library-routines, hereby
reducing the number of load/store operations as well as minimizing 
the number of cache misses.

\begin{figure}[tb]
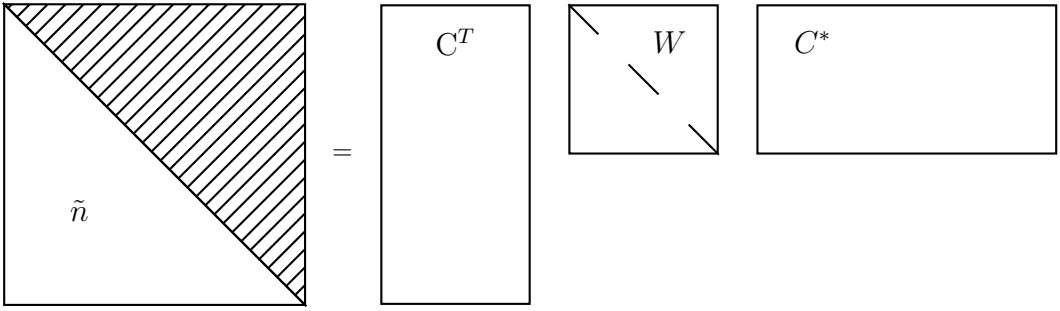

\pspicture(12,4.4)(25.189,9.4)
\pspolygon(12,5)(16,5)(12,9)
\pspolygon[fillstyle=hlines](16,9)(16,5)(12,9)
\psframe(17,5)(19,9)
\psframe(19.5,7)(21.5,9)
\psline[linestyle=dashed,dash=15pt 15pt](19.5,9)(21.5,7)
\psframe(22,9)(26,7)
\rput[c](16.5,7){=}
\rput[c](18,8.5){\large C$^T$}
\rput[c](20.85,8.5){\large $W$}
\rput[c](22.75,8.5){\large $C^*$}
\rput[c](13,6.25){\large $\tilde{n}$}
\endpspicture
\caption{The calculation of the interstitial electron-density in
   \k-space can be regarded as matrix-matrix-multiplication
   $\tilde{n}=C^TWC^*$, where $W$ consists of $j$ identical
   vectors $W_\k(j)$. }
\label{mm_mult}
\end{figure}

\subsubsection{The Interstitial Electron Density}
\nopagebreak
The valence electron density in the interstitial region is given by:

\begin{eqnarray}
n^{IR}({\bf r})&=&\sum_{|{\K}|\leq K^{\rm pot}} n_{{\K}}^{\rm eff}
 (\r) \, \exp(i\K\cdot\r),
\quad \r \in {\rm IR} \\
&=&\sum_{\k,j}\sum_{\bf GG'}W_\k(j) c_\k(j,\G) c^*_\k(j,\G')
\exp(i(\G-\G')\cdot\r)
\end{eqnarray}

where the sum over the occupied states $j$ can again be regarded as 
matrix-matrix-multiplication (see Fig.~\ref{mm_mult}), in which the matrix
$W$ consists of $j$ identical columns $W_\k(j)$:

\begin{eqnarray}
\tilde{n}_\k(\G,\G')&:=&
\sum_{j}\underbrace{W_\k(j) c^T_\k(\G,j)}_{\tilde{c}_\k(j,\G)}
c^*_\k(j,\G')\\
                &=&\sum_{j} \tilde{c}^T_\k(\G,j) c^*_\k(j,\G')
 \label{MATRIX}\\
n({\bf r})&=&\sum_\k\sum_{\bf GG'} \tilde{n}_\k(\G,\G')
\exp(i(\G-\G')\cdot\r)
\end{eqnarray}

Since $W_j(\k)$ is real, the matrix $\tilde{n}({\bf G,G'})$ is
hermitian, {\em i.e.}  $\tilde{n}({\bf G,G'})=\tilde{n}^*({\bf
G',G})$.  Therefore the calculation of the matrix

\begin{eqnarray}
 \tilde{n}_\k({\bf G,G'})=\sum_{j} c^T_\k(\G,j) c^*_\k(j,{\bf G'})
\end{eqnarray}

by a single matrix-matrix-multiplication would result in twice as much 
floating-point operations as necessary, which would destroy the advantage
of using optimized library routines.

To profit from both, the hermiticity of the matrix and the use
of optimized {\sl BLAS-3} library routines, the matrix is divided into
small blocks (Fig.~\ref{blocking}).  Each block above the diagonal is
evaluated by a single {\sl (BLAS-3)} matrix-matrix-multiplication
according to Eq.~\ref{MATRIX} and the result is also used for the
corresponding block below the diagonal.  The elements of the blocks
along the diagonal are evaluated by a direct
implementation of the summation:

\begin{eqnarray}
\tilde{n}_\k({\bf G,G'})=\tilde{n}^*_\k({\bf G',G})=
\sum_{j} c^T_\k({\bf G},j) c^*_\k(j,{\bf
  G'}), \hspace{1cm} {\bf G}\leq {\bf G'}
\end{eqnarray}

\vspace{1cm}

\begin{figure}[hbt]

\pspicture(0,0   )(15.189,7)
\rput(0,-3.5){
{\psset{xunit=.5cm,yunit=.5cm,linewidth=2pt}}

\rput(4,4){

\psframe[linewidth=1pt,linecolor=black,fillstyle=solid,fillcolor=gray](0,4.8)(1.2,6)
\psframe[linewidth=1pt,linecolor=black,fillstyle=solid,fillcolor=white](1.2,4.8)(2.4,6)
\psframe[linewidth=1pt,linecolor=black,fillstyle=solid,fillcolor=white](2.4,4.8)(3.6,6)
\psframe[linewidth=1pt,linecolor=black,fillstyle=solid,fillcolor=white](3.6,4.8)(4.8,6)
\psframe[linewidth=1pt,linecolor=black,fillstyle=solid,fillcolor=white](4.8,4.8)(6,6)

\psframe[linewidth=1pt,linecolor=black,fillstyle=solid,fillcolor=white](0,4.8)(1.2,3.6)
\psframe[linecolor=black,fillstyle=solid,fillcolor=gray](1.2,4.8)(2.4,3.6)
\psframe[linewidth=1pt,linecolor=black,fillstyle=solid,fillcolor=white](2.4,4.8)(3.6,3.6)
\psframe[linewidth=1pt,linecolor=black,fillstyle=solid,fillcolor=white](3.6,4.8)(4.8,3.6)
\psframe[linewidth=1pt,linecolor=black,fillstyle=solid,fillcolor=white](4.8,4.8)(6,3.6)

\psframe[linewidth=1pt,linecolor=black,fillstyle=solid,fillcolor=white](0,2.4)(1.2,3.6)
\psframe[linewidth=1pt,linecolor=black,fillstyle=solid,fillcolor=white](1.2,2.4)(2.4,3.6)
\psframe[linewidth=1pt,linecolor=black,fillstyle=solid,fillcolor=gray](2.4,3.6)(3.6,2.4)
\psframe[linewidth=1pt,linecolor=black,fillstyle=solid,fillcolor=white](3.6,3.6)(4.8,2.4)
\psframe[linewidth=1pt,linecolor=black,fillstyle=solid,fillcolor=white](4.8,3.6)(6,2.4)

\psframe[linewidth=1pt,linecolor=black,fillstyle=solid,fillcolor=white](0,1.2)(1.2,2.4)
\psframe[linewidth=1pt,linecolor=black,fillstyle=solid,fillcolor=white](1.2,1.2)(2.4,2.4)
\psframe[linewidth=1pt,linecolor=black,fillstyle=solid,fillcolor=white](2.4,1.2)(3.6,2.4)
\psframe[linecolor=black,fillstyle=solid,fillcolor=gray](3.6,2.4)(4.8,1.2)
\psframe[linewidth=1pt,linecolor=black,fillstyle=solid,fillcolor=white](4.8,2.4)(6,1.2)

\psframe[linewidth=1pt,linecolor=black,fillstyle=solid,fillcolor=white](0,0)(1.2,1.2)
\psframe[linewidth=1pt,linecolor=black,fillstyle=solid,fillcolor=white](1.2,0)(2.4,1.2)
\psframe[linewidth=1pt,linecolor=black,fillstyle=solid,fillcolor=white](2.4,0)(3.6,1.2)
\psframe[linewidth=1pt,linecolor=black,fillstyle=solid,fillcolor=white](3.6,0)(4.8,1.2)
\psframe[linewidth=1pt,linecolor=black,fillstyle=solid,fillcolor=gray](4.8,1.2)(6.0,0)

}
\psline[linecolor=white,linestyle=dashed,dash=15pt 15pt](4,10)(10,4)
\rput[c](4.6,9.35){\white 1,1}
\rput[c](5.8,9.35){\black 1,2}
\rput[c](7.0,9.35){\black 1,3}
\rput[c](8.2,9.35){\black 1,4}
\rput[c](9.4,9.35){\black 1,5}
\rput[c](4.6,8.15){\black 2,1}
\rput[c](4.6,6.95){\black 3,1}
\rput[c](4.6,5.75){\black 4,1}
\rput[c](4.6,4.55){\black 5,1}
}
\endpspicture
\caption{The hermitian matrix  $\tilde{n}_\k({\bf G,G'})$ is divided into 
small blocks. Each of the blocks is calculated by a 
matrix-matrix-multiplication.}
\label{blocking}
\end{figure}

The blocksize is a free parameter which has to be optimized according
to the cache size of the specific platform.

\subsection{{\tt LAPW1}: Setup and diagonalization of the eigenvalue problem}
\label{lapw1}

\subsubsection{Setup of $H$ and $S$}

According to Eq.~\ref{KSexp} the KS eigenstates
are characterized by a set of expansion coefficients
$c_i(\k+\G)~\{i=1,\dots,N_{\rm s}\}$, where $N_{\rm s}$ are the number
of eigenstates to be calculated.  In the following, these expansion
coefficients are viewed as eigenvectors (of length $N_{\rm pw}$) of
the generalized eigenvalue problem
\begin{eqnarray}
\left(H - \epsilon_i S \right) c_i = 0
\label{diag}
\end{eqnarray}
where $H$ is the Hamiltonian and $S$ the overlap matrix. The elements
of $H$ and $S$ are given by
\begin{eqnarray} 
H_{ij} & = & \langle \phi_i  | H | \phi_j \rangle 
\label{matrixel1} \\
S_{ij} & = & \langle \phi_i      | \phi_j \rangle
\label{matrixel2}
\end{eqnarray}
where $ \phi_j$ are the LAPW bsis functions. 
As discussed earlier, one of the main ideas of the FP-LAPW method is
to construct sophisticated basis functions $\varphi$ which provide a good 
approximation to the true wave function $\psi$, so that the number of 
basis functions $N_{\rm pw}$ required to expand $\psi$ with
reasonable accuracy, is kept small. The main drawback of this approach
is that the evaluation of Eq.~(\ref{matrixel1}-\ref{matrixel2}) is
quite demanding. A simple way to reduce the computational effort in 
setting-up $H$  is to consider in the first half of the self consistency
cycle only the spherical average of the potential
({\em i.e.} the LM=(0,0) component). 
Furthermore a considerable speedup on {\sl IBM RS/6000} nodes
was obtained by using an IBM specific mathematical library ~\cite{mass2.4} which
allows a much faster evaluation of trigonometric
functions that are required in Eq.~(\ref{matrixel1}-\ref{matrixel2}).
These subroutines computes the trigonometric functions for a vector of
arguments, hereby minimizing the computational 
costs compared to the serial evaluation of all vector elements.

A combination of these procedures can significantly speed up the generation
of $H$ and $S$ matrices.

\subsubsection{Solving the eigenvalue problem}

As noted before, the standard diagonalization routines could not be improved
significantly on {\sl IBM RS/6000} nodes, since the modifyed LAPACK routines 
together with IBM´s highly optimized scientific ESSL library yields already almost 
optimal performance.
On other hardware platforms (e.g. SGI Power Challenge, DEC-Alpha, Intel PII)
with slower memory bandwidth we could achieve 
a speedup of the diagonalization by more than a factor of two by modifying the
standard LAPACK routines using a hierarchical 
blocking scheme as described in~\cite{Kvas99}.

\subsubsection{Iterative diagonalization}

In contrast to the LAPW method, the plane wave basis set used in the 
PPW method allows an
easy evaluation of Eq.~(\ref{matrixel1}-\ref{matrixel2}), but the
number of expansion functions is much larger. For this reason the
approach to an iterative matrix diagonalization described below is
somewhat different from the one usually adopted in the PPW method.

We implemented two schemes of iterative matrix diagonalization, namely
the Block-Davidson and the Lanczos algorithm. As both methods are
fairly well known, here only general aspects will be discussed, as far
as they concern the FP-LAPW method. For a detailed discussion see
e.g.~\cite{itdiag}.  

Since the KS equation must be solved self-consistently, the matrix $C$ of the
eigenvectors $c_j$ in Eq.~(\ref{diag}) 
are always available (with the exception of the first cycle) from a previous 
cycle, $C^{\rm old}$. Therefore $C^{\rm old}$ can be used to obtain an 
approximate solution to
Eq.~(\ref{diag}).  If $H^{\rm new}$ ($S^{\rm new}$) is the Hamiltonian
(overlap) matrix of the present iteration, then Eq.~(\ref{diag})
can be transformed into the space spanned by the old
eigenvectors. This would be no approximation to Eq.~(\ref{diag}) if
one would include all eigenvectors, because the old and new
eigenvectors span the same space. In practice, however, the
number of calculated states, $N_{\rm s}$, is much smaller (by almost an order 
of magnitude) than the
matrix size, $N_{\rm pw}$. If one would choose $N_{\rm s}$ equal to the
number of occupied states in the solid, $N_{\rm occ}$ the new eigenvectors
would not be improved at all, since the new eigenvectors, $C^{\rm new}$,
would simply be a linear combination of the old eigenvectors.  Here,
we take $N_{\rm s}~=~2N_{\rm occ}$ which was found to be a good
compromise between accuracy and numerical effort.

The old eigenvectors are now viewed as an unitary transformation
\begin{eqnarray}
{C^{\rm old}}^{\dagger} C^{\rm old} & =  & S \quad .
\label{uni}
\end{eqnarray}
In the case $S$~=~$E$ (no overlap, $E$ unity matrix) Eq.~(\ref{uni})
always holds. In the general case Eq.~(\ref{uni}) is only valid,
if $S^{\rm new}~\simeq~S^{\rm old}$. This aspect must be especially 
considered in the case of the LAPW method because the basis functions
are recalculated in each iteration. This problem can be overcome by
transforming the generalized eigenvalue problem to a regular one ({\em i.e.} by
Cholesky decomposition).  Here, we chose to treat the generalized
problem with the Block-Davidson scheme and the
regular problem using the Lanczos algorithm. The reason for
this strategy is the following: The Lanczos algorithm has due to its
simplicity a very low numerical cost and thus can compensate for
the extra cost of the Cholesky decomposition. Treating the overlap matrix
$S$ explicitly would require to orthogonalize the sets $H^i B^{i-1}$
using $S$ as a metric tensor which would ruin the numerical effort
saved by not doing the decomposition.

The reduced eigenvalue problem is then given by
\begin{eqnarray}
\tilde{H} & = & \epsilon \tilde{S} \tilde{C}^{\rm new} \quad {\rm with} 
\label{sub_it1} \\
\tilde{H} & = & {C^{\rm old}}^{\dagger} H C^{\rm old}  \quad ,
\label{sub_it2} \\
\tilde{S} & = & {C^{\rm old}}^{\dagger} S C^{\rm old} \quad {\rm and} 
\label{sub_it3} \\
\tilde{C}^{\rm new} & = & S {C^{\rm old}}^{\dagger} C^{\rm new} \quad .
\label{new_ev}
\end{eqnarray}

The process of iterating the solution of Eq.~(\ref{sub_it1}) consists
of optimizing the $N_s$ basis functions initially given by $\tilde{C}^{\rm new}$ by
adding $N_s - N_{occ} $ linear independent vectors to this set. In the subsequent
discussion the set $\tilde{C}^{\rm new}$ consisting of $N_{\rm s}$
basisvectors will be named $B^0$ and the set of $N_{\rm s}$
basisvectors added in iteration $i$, $B^i$. The actual iteration
procedure then consists of using $\{B^0$,\dots,$B^i\}$ in
Eq.~(\ref{sub_it1}-\ref{sub_it3}), to construct $B^{i+1}$ from
$\{B^0$,\dots,$B^i\}$ and turn back to
Eq.~(\ref{sub_it1}-\ref{sub_it3}). At the end of the iteration
process the eigenvectors are obtained from Eq.~(\ref{new_ev}).  Here,
the set $\{B^0$,\dots,$B^i\}$ is viewed at as a rectangular matrix of size
$N_{\rm pw}\times(i+1)N_{\rm s}$.

\subsubsection{Lanczos Scheme}\label{secL}
As already mentioned above, we now take $S$~=~$E$.  The basic idea is
to improve $B^i$ by the $N_{\rm s}$ vectors obtained from calculating
$H B^{i-1}$ and orthogonalizing this set to the set $B^{i-1}$ (e.g. by
Gram-Schmidt orthogonalization). In fact this is one of the easiest
ways to increase the basis set, because in practise $H B^{i-1}$ had to
be calculated already in Eq.~(\ref{sub_it2}). To our knowledge, a
strict mathematical proof that the series $H B, H^2 B,\dots,H^n B$ should converge
to the eigenvectors of $H$ does not exist, but experience has shown
that this approach is fairly stable and accurate.
\subsubsection{Block-Davidson Scheme}\label{secBD}
This scheme uses a more subtle way to expand the basis $B$.  In
iteration $i$ one gets from Eq.~(\ref{sub_it1}-\ref{new_ev}) a current
approximation to the true eigenvector $| c_j \rangle$, denoted as
$|c^i_j \rangle$.  The aim is to find a correction vector $|\delta
A\rangle$ such that
\begin{eqnarray}
|c_j\rangle & = & |c_j^i\rangle + |\delta A \rangle \quad .
\label{vec_incr}
\end{eqnarray}
This correction vector $|\delta A\rangle$ can be formally
calculated by plugging Eq.~(\ref{vec_incr}) into Eq.~(\ref{diag}).
\begin{eqnarray}
\left( H - \epsilon_j S \right) | c^i_j \rangle & = & 
\left( H - \epsilon_j S \right) |\delta A_j \rangle
\label{ins_diag}
\end{eqnarray}
The left side of Eq.~(\ref{ins_diag}) is called residual vector, $|R_j
\rangle$. In principle the inversion of $( H - \epsilon S)$ would
yield the correct $|\delta A\rangle$, but in practice this is never
done because the computational cost of this inversion would already be
comparable to an exact diagonalization.  Thus, one only retains the
diagonal elements of $( H - \epsilon S)$ to make the inversion 
trivial. Eq.~(\ref{ins_diag}) is then expressed with help of the basis
$B^i$
\begin{eqnarray}
|\delta A_j \rangle & = & \sum_k \frac{\langle b^i_k | R_j \rangle}
{\langle b^i_k | H - \epsilon_j S  | b^i_k \rangle} | b^i_k \rangle
\label{delta_a}
\end{eqnarray}
The matrix containing the $|\delta A_1\rangle,\dots ,|\delta A_{N_{\rm
s}}\rangle$ is then used to increase the basis $B^i$ to $B^{i+1}$.

\section{Examples}
\label{results}
In the following we demonstrate the effect of our improvements on
a huge example, namely a 9 layer slab of (4$\times$2)-Cu\,(110) ({\em i.e.} 72
atoms, 792 valence electrons).  We will compare the CPU-time needed to
reach selfconsistency using our improved code with the original code. 
Additionally we will compare our LAPW code with a
most efficient implementation of the PPW-method \cite{fhi96md}.  The
Cu(110) surface is modelled by a nine layer slab repeated periodically
in all three dimensions and separated by a vacuum zone equivalent to
five substrate layers.  We use a lattice constant of 6.64 bohr, which
corresponds to the theoretical LDA bulk value.
Since both methods scale almost linearly with the number of ${\bf
k}$-points, only one point in the surface
BZ has been used for these benchmarks.  The MT radii are chosen to be
2.20\,bohr.  The kinetic-energy cutoff for the plane wave basis needed
for the interstitial region is set to 13.22\,Ry which leads to
matrix-sizes of the hamiltonian matrix of about 7000$\times$7000. The
partial wave 
$(l,m)$ representation (inside the MTs) is taken up to $l_{\rm max}=
10$.  A plane-wave cutoff energy of 81\,Ry~for the Fourier representation of
the potential is used.  The maximum angular momentum in the $(L,M)$
expansion of the potential inside the atomic spheres is set to $L_{\rm max}= 4$.  
In the PPW
calculations, plane waves up to a kinetic energy of 70 Ry had to be
used, to reach a comparable level of accuracy, but we also include the
CPU-time required for a PPW calculation at 40 Ry.

\subsection{{\tt LAPW2}: Generating the electron density}

The original code {\tt WIEN95} needed 19680 CPU seconds (5h 28m) for
the generation of the electron density on an {\sl IBM RS/6000} node
(Table \ref{lapw2_time}).  The calculation of the electron density
inside the spheres took 8640 CPU seconds (2h 24 m) (44\%), while
11040 CPU seconds (3h 4m) (56\%) were needed for the interstitial
electron density. On this latter part our improvements led to a reduction of
the necessary CPU-time to 322 CPU seconds (5m 22s), which is
equivalent to a speed-up factor of 34.  In the part generating the electron
density inside the MT-spheres, the improvement is not as big, but
still a speed-up factor of 12 could be reached, reducing the CPU-time
to 740 seconds (12 m). The relative weight of the two tasks is
shifted by the optimization to 70\% for the spheres and 30\% for the
interstitial.  With these improvement, the contribution of {\tt lapw2}
to the overall runtime becomes negligible, and thus all further
considerations should focus on the program {\tt lapw1} and it's most
time consuming part, the diagonalization of the Hamilton-Matrix.

\begin{table}[htb]
\begin{center}
\begin{tabular}{|l||r|r||r|r||r|}
\hline
&\multicolumn{2}{c||}{WIEN95}&\multicolumn{2}{c||}{optimized}&\\
\hline
&CPU-time &\% &CPU-time &\%&${\rm speed-up}\atop{\rm factor}$\\
\hline
Spheres       &  2h 24m & 44 & 12m 20s  & 69 & 12 \\
Interstitial  &  3h 4m  & 56 &  5m 22s  & 31 & 34 \\
\hline
Total         &  5h 28m &    & 17m 42s  &    & 18.5 \\
\hline
\end{tabular}
\end{center}
\caption{Distribution of  CPU-time needed for the
different parts of the generation of the new electron density ({\tt lapw2})
comparing the original
version (``WIEN95'') with the new one (``optimized''). The column
(``speed-up factor'') lists the speed-up reached.}
\label{lapw2_time}
\end{table}

\subsection{{\tt LAPW1}}
\begin{figure}[hbt]
\begin{center}
\pspicture(10,11)
\rput[c](4,5){
\psfig{figure=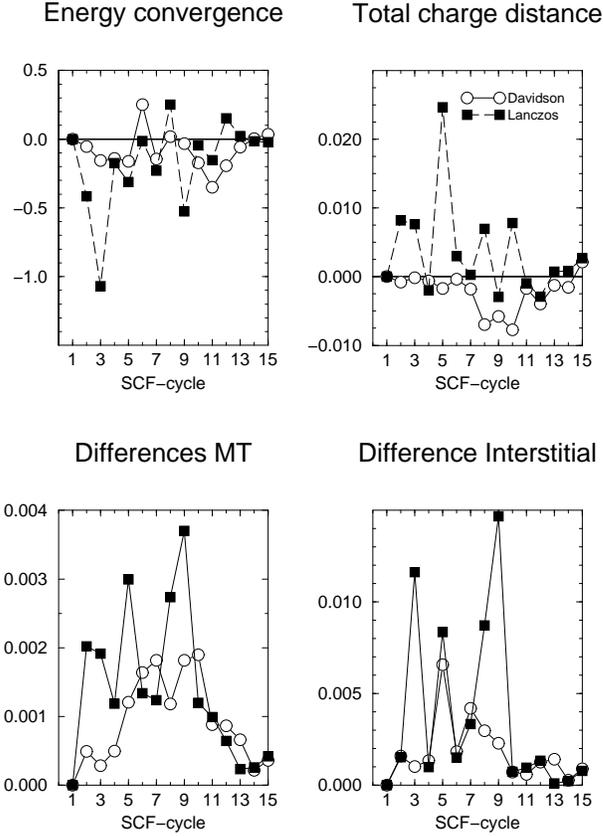,height=12cm}
}
\endpspicture
\end{center}
\label{itpic}
\caption{Upper panel: Deviations of total energy (left panel) in mRy
   and electron distance (right panel) with respect to the exact
   diagonalization for the Lanczos (filled boxes) and Block-Davidson
   (open circles) method during the SCF-cycle.  Lower panel: Valence
   electron distance to the exact solution (see text) for the
   MT-contributions (left panel) and plane wave contributions in
   interstitial region (right panel).  At the start of the SCF-cycle
   an exact diagonalization is performed (no deviation) to obtain the
   input wavefunctions for the Lanczos- and Davidson-method,
   respectively.}
\end{figure}

As a general result it was found that in order to obtain reasonable
accuracy in total energies it is sufficient for both methods, the
Block-Davidson as well as the Lanczos scheme, to improve the expansion
set only once, {\em i.e.} using $\{B^0,B^1\}$ to construct the new
eigenvectors.  Both iterative schemes worked well for the Cu\,(110)
benchmark system.  The speed-up gained with respect to the full
diagonalization was 1.45 in the case of the Lanczos scheme and 3.12 in
the case of the Block-Davidson scheme (Table
\ref{lapw1_lanczos_david}).  Fig.~\ref{itpic} illustrates the accuracy
of both methods during the SCF-cycle.  In the upper panel the overall
performance is illustrated: The left panel shows the deviation of the
total energies obtained by both methods with respect to the exact
diagonalization result. Here, the largest deviation is about 1~mRy,
but when self-consistency is approached, the deviations are well below
the convergence criterion of 0.5~mRy.  The right panel shows in an
analogous way the deviations of the electron differences (the mean
square deviation of $n_{\rm old}-n_{\rm new}$ inside the MT's) during
the SCF-cycle. This gives an idea about the overall quality of the
approximated eigenvectors. The deviation in the total energies are
less than 0.3~mRy and thus show no essential difference between the
two schemes, but the electron difference indicates that the
Block-Davidson method leads to better eigenvectors especially during
the first four cycles of the SCF-cycle.  The quality of the
eigenvectors is illustrated in more detail in the lower panel of
Fig.~\ref{itpic} for the valence electron densities only.  In the left
(right) panel the mean square deviation between $n_{\rm exact}-n_{\rm
iter}$ is evaluated inside the MT's (interstital region), where
``iter'' stands for either the Davidson or the Lanczos method. It can
be clearly seen that the Davidson method leads to results that are
closer to the exact solution than the results obtained by the
Lanczos-method, but again, when self-consistency is reached, both
methods give essentially the same results for the interstitial region
as well as inside the MT.

\begin{table}[hbt]
\begin{center}
\begin{tabular}{|l||r||r||r|}
\hline
& WIEN95 & optimized &\\
\hline
& CPU-time & CPU-time & ${\rm speed-up}\atop{\rm factor}$\\
\hline
spherical $(H, S)$   & 43m 17s     &      16m 29s   & 2.62 \\
non-spherical $(H)$  & 37m 13s     &      18m 36s   & 2.00 \\
Diagonalization      & 1h 12m 5s   & Lan: 49m 42s   & 1.45 \\
Diagonalization      &             & Dav: 23m 07s  & 3.12 \\
\hline
            &                 &  Lan: 1h 24m 47s   & 1.80  \\
\rb{Total}  & \rb{2h 32m 35s} &  Dav:    57m 12s   & 2.67  \\
\hline
\end{tabular}
\end{center}
\caption{CPU-time in LAPW1 needed for the setup (spherical and
   non-spherical H and S matrix) and the diagonalization; for the
   original code the standard diagonalization is used (``WIEN95''),
   while for the ``optimized'' version the timing for both, the
   Lanczos (``Lan'') and the Block-Davidson (``Dav'') method are given
   and the non-spherical part of the Hamiltonian (which is ignored for
   the first half of the iterations towards self-consistency) is the
   average over all iterations.  The last column lists the
   corresponding speed-up factors.}
\label{lapw1_lanczos_david}
\end{table}

\subsection{Total Speed-ups}
\begin{table}[t]
\begin{center}
\begin{tabular}{|l||r|r||r|r||r|}
\hline
&\multicolumn{2}{|c||}{WIEN95}&\multicolumn{2}{c||}{optimized}&\\
\hline
&CPU-time&\% &CPU-time&\%&${\rm speed-up}\atop{\rm factor}$\\
\hline
lapw0 & 26m    &  5  & 26m    & 22 &  1.00 \\
lapw1 & 2h 33m & 30  & 57m    & 61 &  2.67 \\
lapw2 & 5h 28m & 65  & 17m    & 15 & 19.29 \\
core  &   4s   &     &  4s    &    & \\
mixer &   2m   &     &  2m    &  2 & \\
\hline
Total & 8h 29m && 1h 46m &&  4.80\\
\hline
\end{tabular}
\end{center}
\caption{Distribution of CPU-time needed for the different parts of
  the LAPW-selfconsistency cycle comparing the original version
  (``WIEN95'') with the new one (``optimized''). The last column
  shows the speed-up factor reached.}
\label{lapw_time}
\end{table}

Table \ref{lapw_time} shows the distributions of CPU-time needed for
the different parts of the LAPW-selfconsistency cycle for the original
WIEN95 program as well as for our new, optimized code. The enormous
speed-up factor close to 20 for the program {\tt lapw2} indicates,
that the original code, which was tuned for a vector machine, did not
match the needs of modern high performance workstations with fast but
small cache and relativly slow main-memory access.  This extraordinary
speed-up could not be gained for the program {\tt lapw1}.  However,
with the implementation of the new iterative diagonalization
algorithms and the omission of the non-spherical terms to the
Hamilton-matrix in the first half of the selfconsistency run, the
required CPU-time is cut down by a factor of 2.67.  In total all our
modifications lead to a speed-up of 4.80.

\subsection{Comparing the computationally costs
FP-LAPW and Pseudopotentials Plane Waves codes}

In order to compare our improved FP-LAPW-code with the PPW-approach, we also
calculated our test system with the highly optimized PPW code {\tt
fhi96md} \cite{fhi96md}.

In this implementation of the PPW method, the Kohn-Sham-equations are
solved by an iterative optimization of a set of trial wave functions,
combining self-consistency and iterative matrix diagonalization, where
the iteration of the wave functions is formulated in terms of 
equations of motion, as proposed by Car and Parinello
\cite{CarParinello}. In the {\tt fhi96md} code a second order equation
is used, which had been suggested by Joannopoulos
\cite{Joannopoulos}. In this scheme, each single step is
computationally much cheaper than a single selfconsistency cycle
within the FP-LAPW-method, but the number of
iterations needed to reach selfconsistency is usually much larger and
crucially depends on the quality of the initial guess for the wave functions.
For this reason the {\tt fhi96md} code employs a {\sl mixed-basis-set}
initialization, which gives starting wave functions of high quality.
For details see Ref.~\cite {fhi96md}.

\begin{table}[b]
\begin{center}
\begin{tabular}{|l||r|r||r|r|}
\hline
                        & \multicolumn{2}{c||}{FP-LAPW}   & \mc{PPW}        \\     
\cline{1-5}
                        & \multicolumn{1}{c|}{ original}
                        &\multicolumn{1}{c||}{ optimized}&
\multicolumn{1}{c|}{ 70 Ry}&\multicolumn{1}{c|}{ 40 Ry} \\
\hline
$T_{\rm initialization}$  &    30m &    30m  & 18h 40m  & 8h 45m \\
$T_{\rm iteration}$       & 8h 24m & 1h 46m  &  1h  7m  & 30m    \\
\#$_{\rm iterations}$     &  20    &  20     & 100      & 100    \\
\hline
$T_{\rm total}$      &  168h 34m   & 35h 50m & 130h 20m & 58h 45m\\
\hline
\end{tabular}
\caption{CPU-time needed on an {\sl IBM RS/6000} node to converge a
  nine layers slab, representig a 2x2 Cu(110) surface cell. Comparison
  of the original {\tt WIEN95} code (``original''), our improved code
  (``optimized'') and the fhi96md pseudopotential plane wave program
  (``PPW'') with two different plane wave cutoffs (``70 Ry'' and ``40
  Ry'').}
\label{PPW:LAPW}
\end{center}
\end{table}

Table \ref{PPW:LAPW} shows the CPU-time needed to converge our test
system using both iterative matrix diagonalization methods. The
initialization in the FP-LAPW-method is just the time needed to
construct a starting electron density, whereas for the PPW-method the
time reflects the set up of the starting wave functions within the
mixed-basis scheme.  As already mentioned, the time needed for a
single iteration is much smaller for the {\tt fhi96md} implementation
of the PPW-approach than in the FP LAPW code, but this advantage is
destroyed by the fact, that about five times as many iterations are
needed to reach selfconsistency.  It should be noted that the meaning
of ``iteration'' is in fact different in the FP-LAPW and the PPW
method, as the PPW method~\cite{fhi96md} combines the iterative
diagonalization with the selfconsistent update of the electron
density.  While the original {\tt WIEN} code needed about 30\% more
cpu-time than the PPW-code to converge this system, our improved
version is about three times faster.

It is important to note that Table~\ref{PPW:LAPW} summarizes
benchmark-calculations performed in summer 1997, and that the system
Cu (110) with a ($2 \time 4$) surface structure and 72 atoms per
supercell was most favorable to identify the advantages of the new
FP-LAPW code, and at the same time it was least favorable for the
plane-wave pseudopotential code fhi96md. In the meantime several
improvements are being introduced in the pseudopotential code, as for
example a real-space projector method \cite{king-smith} to evaluate
the pseudopotential matrix-elements (which brings a speed up between a
factor of 2 and 3), and ultra-soft pseudopotentials \cite{vanderbildt}
(which brings a speed up by another factor of 2). Altogether, for the
chosen benchmark system the new version of the plane-wave
pseudopotential code, fhi99md, is about a factor of 20 faster, without
loss in accuracy \cite{neugebauer}. But we also note that for other
systems the difference in CPU time required for the new, fhi99md, and
the older, fhi96md, code is much less pronounced.

Other plane-wave pseudopotential codes \cite{VASP,CASTEP} also employ
the mentioned improvements and behave similar to the fhi99md code.
This discussion shows that comparisons between different methods
(e.g., FP-LAPW versus plane-wave pseudopotentials) is indeed helpful
to identify and optimize time critical algorithms and routines. With
ever increasing system size program developments are getting more and
more important. Although FP-LAPW was ahead the pseudopotential code
(with respect to lower CPU time consumption) for some systems in 1997
and 1998, recent improvements by introducing new concepts at the
plane-wave pseudopotentials front make this again a more efficient
code. We are convinced, however, that new concepts and techniques will
also bring a speed up to FP-LAPW. Clearly FP-LAPW remains the most
accurate tool and does not suffer from problems as linearization of
core-valence exchange-correlation (which can be {\em partially}
corrected in pseudopotential calculations), or the lack of core
polarization (which may be important, e.g. for some magnetic systems).
However, besides accuracy low CPU time requirements are clearly very
important. A fast (i.e.  efficiently) working electronic structure
code is crucial for present days problems, in particular to be able to
test all relevant numerical approximations with the required care. We
note that in many density-functional theory calculations performed for
low symmetry and/or many-atom systems the main approximations are
(often) not at the level of exchange-correlation functional but at the
level of numerical approximations.

Although our test system may be a special case and other systems or a
different computer architecture may lead to slight modifications, the
fair estimate of the relative speed between FP-LAPW and PPW should
remain valid.

\section{Summary}
The present work demonstrates that a continuous adaption of algorithms
to the existing hardware architecture is indeed very important for
efficient and accurate electronic structure calculations of many-atom
systems. While the {\tt WIEN 95} implementation of the FP-LAPW-method
was optimized for a vector computer and performs well on those
platforms, it is not well suited for modern cache-based processors.
Our improvements led to a significant speed-up on those hardware
achitectures and makes the FP-LAPW method a strong competitior to the
popular PPW approach. Especially for transition metal systems, 
the FP-LAPW method has a significant advantage. In addition
the FP-LAPW method gives as an all-electron method additional
information about the system, which is out of reach for any
pseudopotential method because of the frozen core approximation.

The significant improvements discussed here have been implemented in
the new version WIEN97 of the FP-LAPW code ~\cite{wien97} and the
sucessful strategy adopted here may be useful for other software
developers too.

\section{Acknowledgements}
We thank R.~Reuter of IBM Heidelberg, Germany for his very helpful
introduction into the art of producting high performance computer
codes for numerically intensive computations. Helpful discussion with
P.~Kaeckell, who performed some calculations with the VASP~\cite{VASP}
code, are greatfully acknowledged.  P.~B. and K.~S. were supported in
part by the Austrian Science Foundation (SFB project F1108), M.~P. by
the Deutsche Forschungsgesellschaft, Sonderforschungsbereich 290.
This work was also supported by the TMR network ``Electronic Structure
Calculations of Materials Properties and Processes for Industry and
Basic Sciences''.

\clearpage

{\renewcommand{\,}{$\!$\ }}

\end{document}